# A Reactive Molecular Dynamics Study of Hydrogenation on Diamond Surfaces

Eliezer F. Oliveira[1,2]*, Mahesh R. Neupane[3], Chenxi Li[4], Harikishan Kannan[4], Xiang Zhang[4], Anand B. Puthirath[4], Pankaj B. Shah[3], A. Glen Birdwell[3], Tony G. Ivanov[3], Robert Vajtai[4], Douglas S. Galvao[1,2], Pulickel M. Ajayan[4]*

[1]Group of Organic Solids and New Materials, Gleb Wataghin Institute of Physics, University of Campinas (UNICAMP), Campinas, SP, Brazil.
[2]Center for Computational Engineering & Sciences (CCES), University of Campinas (UNICAMP), Campinas, SP, Brazil.
[3] CCDC US Army Research Laboratory, Adelphi, MD, USA
[4]Department of Materials Science and Nanoengineering, Rice University, Houston, TX 77005, USA

*Email address and phone number: efoliver@ifi.unicamp.br, +55(14)98142-0122; ajayan@rice.edu +1 713-348-5904

Abstract:

Hydrogenated diamond has been regarded as a promising material in electronic device applications, especially in field-effect transistors (FETs). However, the quality of diamond hydrogenation has not yet been established, nor has the specific orientation that would provide the optimum hydrogen coverage. In addition, most theoretical work in the literature use models with 100% hydrogenated diamond surfaces to study electronic properties, which is far from the experimentally observed hydrogen coverage. In this work, we have carried out a detailed study using fully atomistic reactive molecular dynamics (MD) simulations on low indices diamond surfaces i.e. (001), (013), (110), (113) and (111) to evaluate the quality and hydrogenation thresholds on different diamond surfaces and their possible effects on electronic properties. Our simulation results indicate that the 100% surface hydrogenation in these surfaces is hard to achieve because of the steric repulsion between the terminated hydrogen atoms. Among all the considered surfaces, the (001), (110), and (113) surfaces incorporate a larger number of hydrogen atoms and passivate the surface dangling bonds. Our results on hydrogen stability also suggest that these surfaces with optimum hydrogen coverage are robust under extreme conditions and could provide homogeneous p-type surface conductivity in the diamond surfaces, a key requirement for high-field, high-frequency device applications.

1. Introduction:

Recently, it was demonstrated that the use of diamond as a p-type material can be a relevant innovation for the new generation of field-effect transistors (FETs) [1–7]. Diamond has high carrier mobility, high thermal conductivity, and resistant to extreme environmental conditions [8,9]. Besides this, the cost of manufacturing diamonds is decreasing rapidly [8]. These qualities make diamond-based FETs an ideal material for the next generation of electronics that could equip spacecraft and high-frequency devices.

In the diamond-based FETs, it is common to use hydrogenated diamond to passivate the remaining dangling bonds of the exposed carbon atoms at the surfaces and make the surface less reactive (to avoid reactions with the material used as a gate, for example) [10–14]. Also, the hydrogenation of diamond decreases its work function (WF) and results in a negative electron affinity (EA), turning it into a p-type material, suitable to FET applications [8,9,12,13,15,16]. Beyond that, hydrogenation can also be used to create a channel of holes that contributes to the conductivity of charge carriers in the surfaces interfaced with oxide-based acceptor layer with large WF [6,10,11,13,17].

During the diamond crystal syntheses (powder or film), different exposed facets/surfaces can be created [8–10]. These diamond surfaces will have a different number of carbon dangling bonds, structural defects, and surface reconstructions [10,13,18,19]. Such features will have a direct impact on the hydrogenation patterns and on the resulting electronic and mechanical properties. It is well known that the hydrogenated surface layer significantly influences the surface conductivity. In this sense, diamond crystal with the most exposed surface that presents the minimum number of dangling bonds and the maximum number of bonded hydrogen atoms would result in efficient electrons transfer from the hydrogenated diamond surface to the acceptor layer. Consequently, this charge equilibrium process contributes to the formation of a p-type (hole) channel in the diamond surface.

To the best of our knowledge, a comprehensive study on the hydrogenation dynamics/patterns on different diamond surfaces is still lacking. Moreover, most of the reported theoretical studies consider a 100% hydrogenated surface [13,19,20], which is unrealistic from the experimental point of view. In this work, using fully atomistic reactive MD simulations, we addressed some of these points considering scenarios closer to the experimental conditions.

Our results are organized as follows. First, we briefly discuss the behavior of the different bare diamond surfaces to determine the number of existing dangling bonds. The information on the dangling bond densities is then used to estimate the percentage of hydrogen coverage on the surface. Following that, we will discuss the structural of diamond surfaces by comparing their total formation energies. Finally, we will analyze the hydrogenation dynamics/patterns of each diamond surface to estimate the threshold hydrogen limit in terms of how many hydrogen atoms can be incorporated into the structures.

2. <u>Methods:</u>

We used fully atomistic reactive molecular dynamics (FARMD) simulations to evaluate the characteristics of the different diamond surfaces and to determine the effectiveness of the hydrogenation process for each one. All FARMD simulations presented here were performed with the ReaxFF force field [21], as implemented in the computational LAMMPS [22] code. The ReaxFF parameters were obtained from the work of K. Chenoweth *et al*. [23].

We started by selecting five different pristine diamond surfaces: (001), (013), (110), (113), and (111). These surfaces are selected because they are widely studied in several experimental and theoretical works for device applications [6,7,13,18–20,24–26]. In addition, these surfaces are the most "exposed" ones in diamond growth processes using structural seeds [8]. For the bare surfaces, we evaluated the number of available carbon dangling bonds on the surface, identified the probability of structural reconstructions, and analyzed the surface stability. Using the observed physical and structural parameters, we can not only estimate the number of hydrogen atoms that can be incorporated into each surface, but also identify the type of adsorbed/terminated hydrogen functional groups (-CH, -$CH_2$, or -$CH_3$).

To evaluate the characteristics of each diamond surface, we created square diamond slabs of ~5.4 x 5.4 $nm^2$ surface area, with the thickness of eight layers, which is enough to preserves bulk-like properties [13,19]. The positions of the carbon atoms in the bottom two layers were kept constrained in the diamond bulk positions to mimic bulk-like slab and reduce the computational cost. The remaining six layers were allowed to freely move during the simulations. Initially, we performed an energy minimization of each diamond surface using the conjugate gradient technique with an energy convergence tolerance of 0.001 kcal/mol and a force convergence tolerance of 0.5 $kcal.mol^{-1}Å^{-1}$. These energy minimizations were followed by thermal equilibrations to study the surface characteristics and to analyze their structural stability at different temperatures (room temperature (RT), 500, 700, 900, and 1200°C) for 400 ps, which is time enough to allow structural transitions and/or equilibrations. For each case, the temperature was ramped from –273°C to a desired temperature. The duration of the whole process is typically ~500 ps. All simulations were carried out using constant-temperature, constant-volume ensemble (NVT) and a time step of 0.1 fs. The temperature of the simulations was controlled using the Nosé-Hoover thermostat and periodic boundary conditions (PBC) were applied along the x and y directions to reproduce infinite slab. The dimensions are large enough to preclude spurious mirror effects.

After studying the properties of each bare diamond surface, we performed the hydrogenation process simulation on these surfaces. An atmosphere of randomly distributed atomic hydrogen atoms was created, and they were exposed to thermally equilibrated diamond surfaces. We considered an atmosphere with the hydrogen atoms 20% higher than the available dangling bonds. The hydrogen atoms were not allowed to recombine with each other in order to speed up the simulations. Similar to the bare simulations, the PBC along the x and y directions was applied, but the system along the z-direction was confined by adding two infinitely hard Lennard-Jones walls at the bottom and top of the simulation box. Then, for each diamond surface, we carried out FARMD runs for 2.0 ns at 500°C, 700°C, 900°C, and 1200°C using the NVT ensemble. The simulation box size was chosen to keep the internal pressure at 1 atm. The time-step used in these simulations was 0.1 fs, which is similar to the one used in the bare surface simulations. Using the results from these simulations, the effectiveness of each diamond surface, with respect to the number of incorporated hydrogen atoms as a function of the simulation time, can be evaluated and analyzed. The hydrogenation procedure adopted here has been already used with success in similar studies with other carbon-based materials [27–31].

3. Results and discussion:

3.1. Bare surfaces:

3.1.1. Surface (001):

In Figure S1, we present the structural model of the (001)-1x1 diamond surface in its initial configuration before the thermal equilibration. The carbon atoms highlighted in red have initially two dangling bonds, which is equivalent to ~32 dangling bonds per $nm^2$. These surface carbon atoms are at a distance of ~2.5 Å from each other.

Initially, when the temperature of the slab was ramped up, the bulk-like surface initiates the surface reconstruction process, as shown in Figure 1(a). As can be seen, at the thermal equilibration of 900°C, some of the most exposed/surface carbon atoms (highlighted in red) form bonds with their neighbors, decreasing the number of dangling bonds and thus generating a reconstruction known as 2x1. This is in good agreement with the experimental data and with results from electronic structure simulations [10,13,19,20]. Unlike earlier theoretical observations, our results suggest that the reconstruction is temperature dependent. We observed that the degree of reconstruction varies from 10- 85% when temperature varied from room temperature to 1200°C, respectively. This corresponds to approximately 30, 22, 19, 18, and 18 dangling bonds/$nm^2$ on the surface at RT, 500, 700, 900, and 1200°C, respectively. The optimum value of the dangling bonds at each temperature indicates the most favorable atoms to be hydrogenated.

The carbon bond-lengths in the bulk and surface exhibit some fluctuations, as expected, for different temperatures. In the bulk region, at RT, the C-C bond-length is 1.55 ± 0.03 Å, with slight fluctuation of ~0.02 Å for higher temperatures. As expected, the bond characteristic of these atoms is of $sp^3$-$sp^3$ type. The reconstructed carbon atoms (dimers), as indicated in Figure 1(b) by the bonds among the atoms C1, present a bond length of 1.42 ± 0.02 Å, at RT, assigned as a $sp^2$-$sp^2$ bond. The bond lengths between the carbon atoms (C1 and C2 in Figure 1(b)) are 1.52 ± 0.05 Å, at RT, similar to a $sp^3$-$sp^2$ character. However, increasing the temperature further leads to an increment of ~0.02 Å in the bond lengths. The bond lengths of the bonds among the atoms (C1 and C3 in Figure 1(b)) that do not participate in the 2x1 reconstruction are 1.50 ± 0.05 Å, at RT, and can also be assigned as a $sp^3$-$sp^2$ character. Increasing the temperature further results in an increase of ~0.01 Å in these values. The values for the bond lengths between the carbon atoms at the surface are in agreement with previous studies performed with electronic structure simulations [6,7,13,19,20]. It is important to stress here that these bond-length values were obtained through an average along the total elapsed time of simulations.

### 3.1.2. Surface (013):

In Figure 1(c), we present the structural model of the (013) diamond surface. This phase was first observed by Silva et al. [32] during (100) diamond epitaxy growth and is relatively less widely used. Unlike the (001) surface, this surface did not exhibit any temperature-dependent surface reconstructions, maintaining a constant number of the dangling bonds for all the applied temperatures. The carbon atoms highlighted in red (see Figure 1(c)) have two dangling bonds while the green ones have one. This surface contains ~29 dangling bonds per $nm^2$ and the surface carbon atoms are at a distance of ~3.1 Å from each other. The average C-C bond length in the bulk region is 1.55 ± 0.03 Å, at RT, a $sp^3$-$sp^3$ bond characteristics. At the surface, however, as shown in Figure 1(d), the surface atoms represented by C1 and C2 (the ones with dangling bonds) have bonds with bond lengths of 1.45 ± 0.03 Å, which can be assigned as $sp^2$-$sp^2$ bond. The carbon atoms represented by C1 and C3 in Figure 1(d) possess bond-length of 1.54 ± 0.03 Å, at RT. The bond-length of the atoms C2 and C3 are 1.53 ± 0.03 Å, at RT, little smaller than a formal a $sp^3$-$sp^3$ bond. As expected, increasing the surface temperature results in an elongated bond with an increment of ~0.02 – 0.03 Å. These bond-length values were obtained through an average along the total elapsed time of simulations.

### 3.1.3. Surface (110):

The structural model of the (110) diamond surface is presented in Figure 1(e). This surface also did exhibit any temperature-dependent surface reconstruction, consistent with earlier theoretical observations [13,19]. The carbon atoms highlighted in red (see Figure 1(e)) have one dangling bond each and they are distributed in a zig-zag configuration. This surface possesses ~22 dangling bonds per nm$^2$, and as this surface does not reconstruct, this number remains the same during all the tested temperatures, i.e., from RT up to 1200°C. For this diamond surface, in the bulk part the C-C bonds have an average length of 1.55 ± 0.04 Å at RT, indicating a sp$^3$-sp$^3$ bond. At the surface, as shown in Figure 1(f), the bond length among the carbon atoms represented by C1 is 1.46 ± 0.03 Å, suggesting a sp$^2$-sp$^2$ character. As for the bonds among the C1 and C2 carbon atoms (see Figure 1(f)), we observed bond-length values of 1.53 ± 0.03 Å at room temperature, a little smaller than a sp$^3$-sp$^3$ bond. As for the bonds among the atoms represented by C2 in Figure 1(f), the bond-length value is 1.55 ± 0.04 Å at room temperature, indicating a sp$^3$-sp$^3$ character. As with the other surfaces, increasing the surface temperature results in elongated bonds by ~0.02 – 0.03 Å.

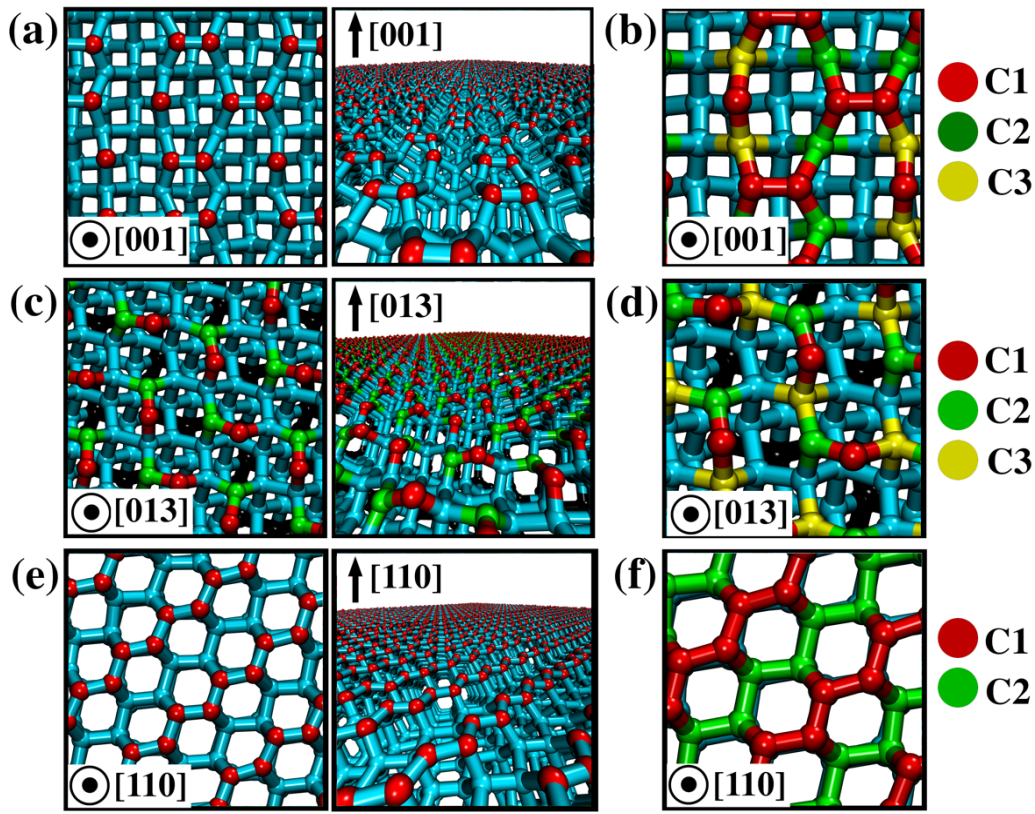

**Figure 1.** (a) Representative MD snapshots of the (001) diamond surface after a thermal equilibration at 900°C in a top and perspective view. The red atoms represent the carbon atoms that have dangling bonds. (b) Highlighted atoms in the (001) surface indicating the different bond lengths between the carbon atoms at the surface (see text for discussions).

(c) Representative MD snapshots of the (013) diamond surface after a thermal equilibration at 900°C in a top and perspective view. The red and green atoms represent the carbon atoms that have dangling bonds. (d) Highlighted atoms in the (013) surface to indicate the different bond lengths between the carbon atoms at the surface (see text for discussions). (e) Representative MD snapshots of the (110) diamond surface after a thermal equilibration at 900°C in a top and perspective view. The red atoms represent the carbon atoms that have dangling bonds. (f) Highlighted atoms in the (110) surface to indicate the different bond lengths between the carbon atoms at the surface (see text for discussions).

### 3.1.4. Surface (111):

The structural model of the (111) diamond surface is illustrated in Figure 2(a). This is the most sought-after phase for quantum applications. In spite of its' applicability in the quantum as well as RF applications [3,33], this phase is the least understood phase in terms of its reconstruction mechanism [7]. At room temperature, we did not observe any structural reconstruction found in (100) phase, but the surface exhibited a quasi-planar configuration. The carbon atoms highlighted in red in Figure 2(a) have one dangling bond each, resulting in ~18 dangling bonds per $nm^2$. However, with the increasing temperature, some regions of the surface undergo a structural transition to 2x1 configurations, as indicated in Figure 2(b). This increases the number of dangling bonds to 19, 21, 22, 23 dangling bonds per $nm^2$ for temperatures of 500, 700, 900, and 1200°C, respectively.

In our MD simulations, we did not observe a complete 2x1 reconstruction for the (111) surface. The configuration shown in Figure 2(b) agrees with the one predicted by Refs [7,34] as a transition state between an unreconstructed 1x1 phase and a fully reconstructed 2x1 phase. According to the literature, the most stable configuration for the (111) surface is a complete 2x1 reconstruction [7,10,18–20]. Since the diamond surfaces used in this study are defect-free, our simulations suggest that other effects might be necessary to trigger a complete 2x1 reconstruction. The hydrogenation of the (111) surface recovers the 1x1 configuration mainly due to the nominal energy barrier between these phases [10]. For this 1x1 (111) surface, the observed bond-length values for the surface carbon atoms represented by C1 and C2 in Figure 2(c) are 1.44 ± 0.02 Å at room temperature, suggesting a $sp^2$-$sp^2$ characteristics. However, the bonds among the carbon atoms represented by C2 and one layer below are of $sp^3$-$sp^3$ characteristics with the bond length of 1.55 ± 0.04 Å at room temperature, a bond-length fluctuation of ± 0.03 Å.

### 3.1.5. Surface (113):

In Figure 2(d) we present the structural model of the (113) diamond surface. The carbon atoms highlighted in red (see Figure 2(d)) have two dangling bonds, while the green ones have just one dangling bond. The red carbon atoms have physical separations of ~2.5 Å from each other. For this surface, the surface reconstructions were not observed up to 900°C, but, at 1200°C, a few reconstruction phases start to occur, as shown in Figure 2(e). Up to 900°C, this surface presents ~30 dangling bonds per $nm^2$, but the density of the dangling bonds decreases slightly to 28 dangling bonds per $nm^2$ when the temperature was increased to 1200°C due to the surface reconstructions. Similar to the other surfaces discussed so far, the average C-C bond-length is 1.55 ± 0.02 Å, at room temperature, in the bulk region, but this average value varies up to ~0.02 Å with the increased temperature. The bond characteristics of these bonds are of the $sp^3$-$sp^3$ types. At the surface, as shown in Figure 2(f), the bond-length of the carbon atoms represented by C1 and C2 (the ones with dangling bonds) is 1.46 ± 0.02 Å, which is an indication of a $sp^2$-$sp^2$ characteristics. The bonds among the carbon atoms represented by C2 and C3 in Figure 2(f) have bond-lengths of 1.56 ± 0.04 Å. For the atoms represented by C1 and C4, the bond-lengths are slightly smaller with 1.53 ± 0.03 Å. Further analysis of the bond-lengths and corresponding angles reveals that the bond characteristics of these bonds are of a $sp^3$-$sp^3$ type. Increasing the surface temperature resulted in an increment in the bond-lengths of the carbon atoms at the surface.

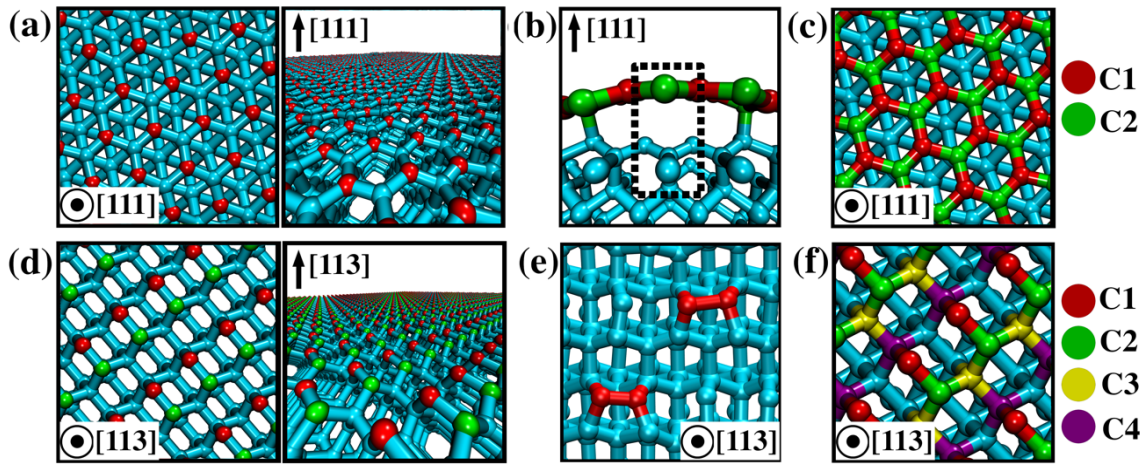

**Figure 2.** (a) Representative MD snapshots of the (111) diamond surface after a thermal equilibration at 900°C in a top and perspective view. The red atoms represent the carbon atoms that have dangling bonds. (b) Representation of the transition states between 1x1 to 2x1 reconstruction for the (111) surface when the temperature is increased. (c) Highlighted atoms in the (111) surface to indicate the bond lengths among the carbon atoms at the surface (see the text for discussions). (d) Representative MD snapshots of the (113) diamond surface after a thermal equilibration at 900°C in a top and perspective view. The red and green atoms represent the carbon atoms that have two and one dangling bond, respectively. (e) Structural reconstructions (indicated in red) occur in the (113) surface when the temperature is above 900°C. (f) Highlighted atoms in the (113) surface to indicate the bond lengths among the carbon atoms at the surface (see the text for discussions).

### 3.1.6. Surface stability:

The stability of hydrogenation is also useful to know for fabricated devices. The surface temperature of hydrogenated diamond transistors can spike within nanoseconds during pulse turn-on, and then experience a more gradual temperature increase within microseconds to seconds. Certain surface orientations would offer better stability during device operation. The surface stability study of each diamond surface studied here can provide helpful insights on their relative feasibility to be produced/synthesized for device applications. In Figure 3, we present the potential energy per atom with the elapsed time of the MD simulations at RT (Figure 3(a)), 500°C, (Figure 3(b)), and 900°C (Figure 3(c)). In this figure, we only considered the potential energy of the non-constrained atoms. At RT, the most stable surface is the (110), followed by the (111), (013), (113), and (001), respectively. For the surface (001), we observed a decrease in the per atom potential energy for a duration of ~55 ps of simulation; this can be attributed to the structural reconstructions that are occurring in the surface—when the reconstructions stop, the potential energy remains essentially constant (after ~55 ps).

When the temperature of each diamond surface is increased, as shown in Figures 3(b, c), the ordering of the most stable surfaces changes to (110), (111), (001), (013), and (113). Compared to the data at room temperature, increasing the temperature, the surface (001) becomes more stable than the (013) and (113) surfaces. This is mainly due to the increased level of structural reconstructions, which in turn stabilizes the surface. As discussed in Section 3.1.1, the probability of surface reconstructions in the surface (001) increases from 10% to 75% when the temperature is increased from RT to 500°C. In summary, these results are in agreement with other theoretical/modeling works with different methodologies. In these studies, the stability was verified by comparing the electronic structures and surface formation energies and confirmed that the surface (110) is the most stable one, followed by the (111) and (001) surfaces [13,19,20]. We did not find in the literature data for the structural stability of the (013) and (113) surfaces.

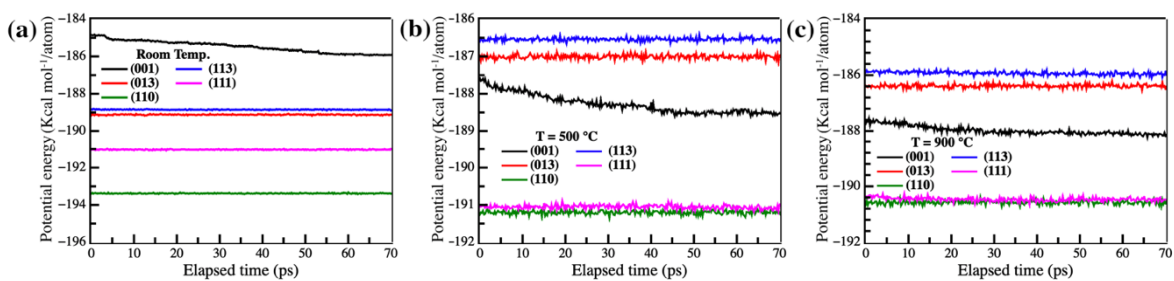

**Figure 3.** Potential energy per atom in relation to the elapsed time of the MD simulations at (a) room temperature, (b) 500°C, and (c) 900°C.

3.1.7. Analyses of dangling bond density:

The degree of hydrogen coverage during the hydrogenation process depends on the number of dangling bonds, i.e., bond density, on the surface. From the MD simulations performed for the bare diamond surfaces (001), (013), (110), (111), and (113), it is possible to estimate the dangling bond densities on these surfaces. Table 1 presents the number of dangling bonds in each surface from the simulations discussed in the previous sections. The (001), (111), and (113) surfaces exhibit slight variations in the bond density due to the surface reconstructions when the temperature is varied from RT to 1200C. The (113) presents changes in the number of dangling bonds only for very high temperatures (1200°C), but the (013) and (110) surfaces are robust against temperature variation.

**Table 1.** The number of dangling bonds per nm² for the different diamond surfaces as a function of the temperature.

| Surface | Temperature | | | | |
|---|---|---|---|---|---|
| | **Room temperature** | **500°C** | **700°C** | **900°C** | **1200°C** |
| **(001)** | 30 | 22 | 19 | 18 | 18 |
| **(013)** | 29 | 29 | 29 | 29 | 29 |
| **(110)** | 22 | 22 | 22 | 22 | 22 |
| **(113)** | 30 | 30 | 30 | 30 | 28 |
| **(111)** | 18 | 19 | 21 | 22 | 23 |

3.2. Hydrogenation of diamond surfaces:

Hydrogenation is a critical surface treatment method implemented to induce dipole-mediated surface charges in insulating diamond surface. After the initial evaluation of the bare diamond surfaces as a function of temperature and estimation of dangling bond densities in all the surfaces, the hydrogenation simulations were performed. The initial diamond surface structures were previously obtained from the thermalization processes at four different temperatures: 500, 700, 900, and 1200°C. As mentioned in the Methodology section, the surfaces were exposed to an atmosphere of hydrogen atoms, and the simulated hydrogenation processes were carried out at these temperatures. The number of hydrogen atoms in the atmosphere is equal to 120% of the number of dangling bonds of the thermalized structures.

In Figure 4(a) we present the MD snapshots of the initial configurations of each tested diamond surface in contact with the atmosphere of hydrogen atoms. Experimentally, the most common temperatures at which the diamond hydrogenation is performed are in the range of 700–900°C, using hydrogen plasma [10,35–41]. In order to gauge how the probability of hydrogenation changes with temperature, the hydrogenation simulation was done for the temperature range between 500-1200 C. In Figure 4(b) we present MD snapshots in perspective and top views of the resulting hydrogenated diamond surfaces at 900°C after the hydrogenation. A better understanding of the whole hydrogenation process at this temperature can be seen in the Supplementary Videos SV1 to SV4. The hydrogenation incorporation occurs randomly without an apparent preference for specific sites to be chemically bonded, as evident from Figure 3(b) and the supplementary videos. This can be ascribed to the charge fluctuations that occur at the surface because of the fluctuating bond lengths and charges over time. The process of C-H bond formation continues until the hydrogen saturation threshold is reached. In addition, when many hydrogen atoms start to accumulate in a specific region, they electrostatically modify their neighborhood, thus decreasing the probability of further hydrogen incorporation.

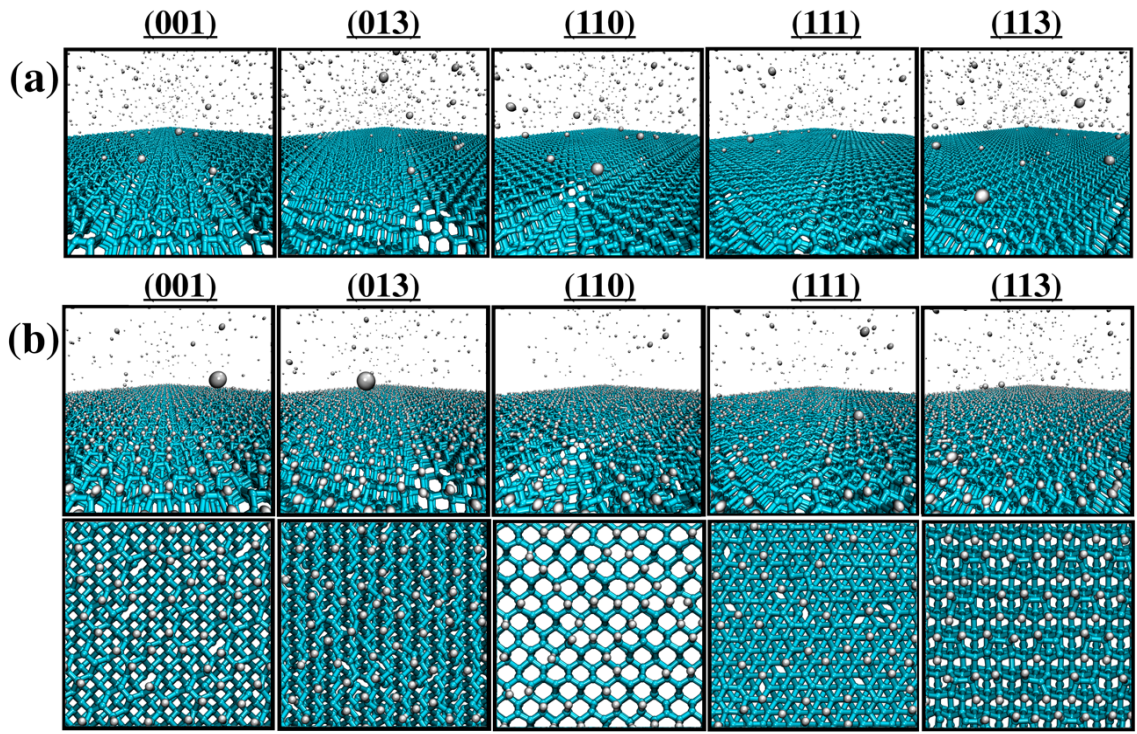

**Figure 4.** (a) Representative MD snapshots of the initial configuration of each tested diamond surface into contact with the atmosphere of hydrogen atoms. (b) Representative MD snapshots of the resulting hydrogenated diamond surfaces at 900°C after the hydrogen incorporation saturation in perspective and top views.

In Figures 5(a) to 5(d) we present the number of hydrogen atoms per nm² that are incorporated in each tested diamond surface as a function of the simulation time for the different temperatures. The data presented in this figure were fitted using the following exponential function:

$$\#H(t) = \#H_o - Ae^{\left(\frac{-t}{B}\right)} \quad (1)$$

In Equation 1, $\#H(t)$ and $\#H_0$ represent the number of hydrogen atoms per nm² that are incorporated into the diamond surface in a time t (in ps) and the saturated one (when the incorporation saturates), respectively, and A and B are adjusted parameters. The data obtained from these exponential fittings are presented in Table S1. The effects of temperature on the hydrogenation dynamics and/or patterns are evident from the plots. All saturation curves suggest that, independent of the temperature and type of surface, at least ~50ps of exposure time is required to reach the hydrogen saturation threshold for each diamond surface.

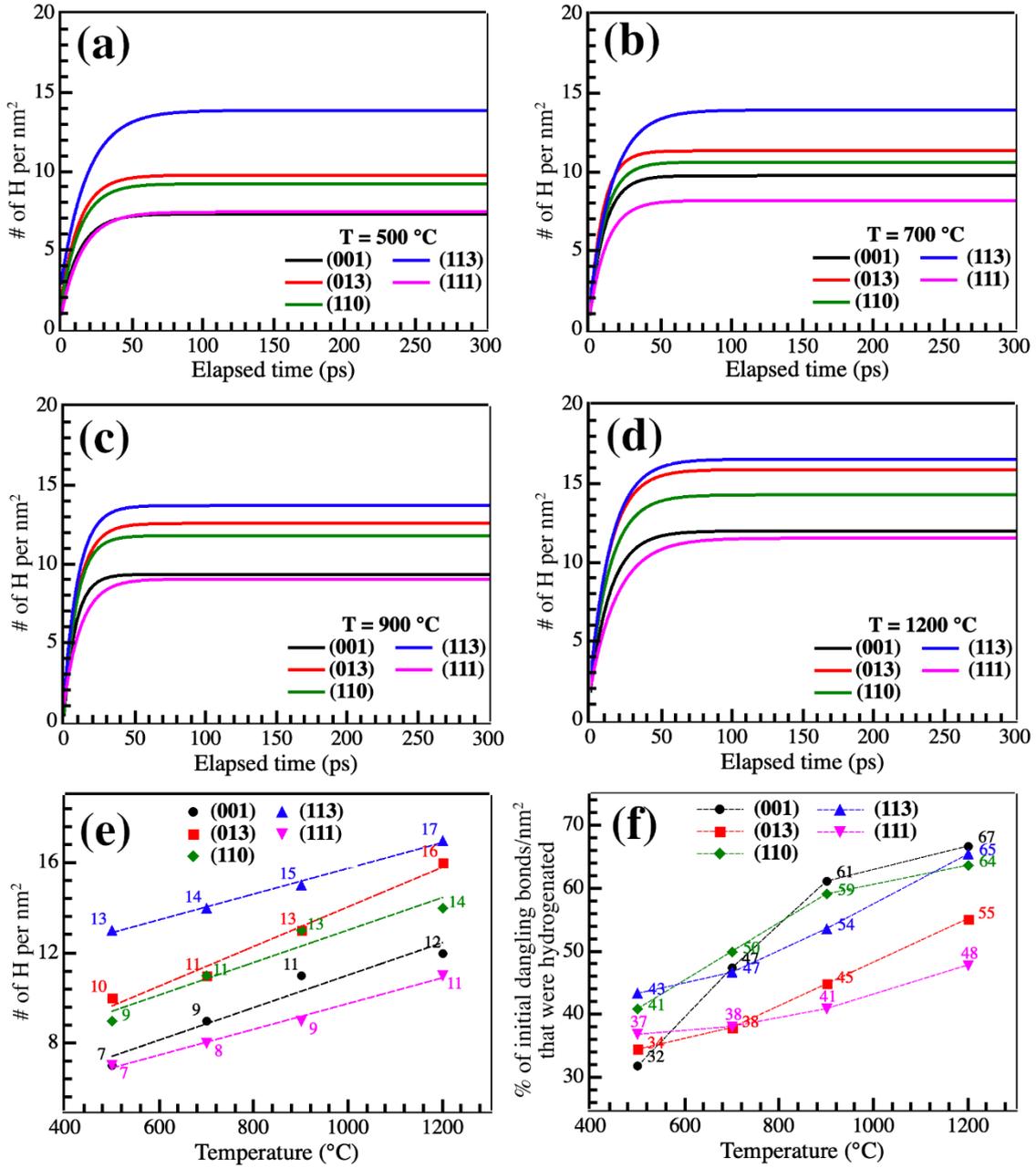

**Figure 5.** Evolution of the number of bonded hydrogen atoms per nm² at different temperatures ((a) 500, (b) 700, (c) 900, and (d) 1200°C) as a function of the simulation time for the investigated diamond surfaces. The data were fitted using Equation 1. (e) The resulting number of hydrogen atoms per nm² after the hydrogenation process saturation threshold is reached for each diamond surface as a function of the temperature. (f) Percentage of the number of initial (pre-hydrogenation) carbon dangling bonds for each diamond surface as a function of the temperature.

From the data obtained for the saturated number of hydrogen atoms incorporated into the diamond surface (parameter $\#H_0$ in the Equation 1 above) for each curve shown in Figures 5(a) to 5(d), we can extract the number of hydrogen atoms that can be incorporated into each surface as a function of the temperature. As can be seen in Figure 5(e), the number of hydrogen atoms incorporated exhibits an almost linear temperature dependence for all studied diamond surfaces. According to the results from the linear fit performed in Figure 5(e) and Table S2, we have approximated one additional hydrogen atom per nm$^2$ in each increase of 140°C, 114°C, 140°C, 175°C, and 175°C for, (001), (013), (110), (113), and (111) surfaces, respectively. Because of the higher dangling bond densities, the greatest number of hydrogen atoms incorporated are for (013) and (113) surfaces (see Table 1). Also, it is important to stress here that we are using pristine (non-defective) diamond surfaces, these numbers can vary in a non-ideal case and/or quality sample. As discussed previously, the diamond hydrogenation results in a negative electron affinity (EA), which allows it to be used as a p-type material in electronic device applications as, for example, in FETs. According to experimental and theoretical data, more hydrogen atoms at the diamond surfaces imply a more negative superficial EA, while concomitantly decreasing the WF [8,9,12,13,15,16,19,20]. If we follow this tendency, then diamond films with the (013) and (113) surfaces with higher probability of hydrogenation could be beneficial to some electronic applications.

However, beyond the number of hydrogen atoms that can be incorporated at the surface, it is also important to verify which surface is most efficient in incorporating hydrogen during the hydrogenation process. Since the bare surface has positive EA, increasing the number of hydrogen atoms results in a reduction in the value of EA down to -1.2 eV, i.e. negative EA, for the fully hydrogenated diamond surface [13,19,42]. Thus, passivating the dangling bonds that still remain at the diamond surface after the hydrogenation process would lead to more negative EA.

In Figure 5(f) the densities of the initial dangling bonds per nm$^2$ that were hydrogenated as a function of the temperature in each surface are presented. These percentages are based on the number of initial dangling bonds for each temperature presented in Table 1 and the total number of hydrogen atoms incorporated into each surface. As can be seen, the percentage of hydrogenation that can be effectively realized can reach 67% at 1200°C. This suggests that it is not possible to reach 100% of surface hydrogenation. This is expected, since increasing the number of hydrogen atoms at the diamond surfaces will also increase the electrostatic interaction (repulsion/steric) among them, which makes a complete hydrogenation impossible. At temperatures lower than 700°C, the surfaces (113) and (110) are the most efficient in passivating the number of dangling bonds, 43% and 41%, respectively. From 700°C up to higher temperatures, the hydrogenation efficiency of the surface (001) increases, and between 700°C and 900°C, this surface becomes the most "hydrogenable", reaching values as high as 60% of the initial dangling bonds. From 900°C to 1200°C, the (001), (110), and (113) surfaces are relatively suitable for the hydrogenation, reaching the highest and similar degree of hydrogenation. The surfaces (013) and (111) appear to be the least efficient ones in passivation. According to these results, even though the diamond surfaces (001) and (110) do not present the highest number of hydrogen atoms at the surface, they are among the best to passivate the dangling bonds during the hydrogenation process at high temperatures. Interestingly, the (113) surface presents both the highest number of hydrogen atoms and considerable passivation at higher temperatures. An experimental study have reported low electrical resistance in boron-doped (113) surface as compared to other phases [8]. Hence, the (113) surface with optimum hydrogen coverage seem to be a very promising candidate for electronic device applications.

Since the (013) and (113) surfaces provide the highest bond densities and hydrogenation, these surfaces warrant additional analysis. As per our earlier discussion, the (013) and (113) surfaces present atoms with two and one dangling bond (the atoms highlighted in red and green, respectively, in Figures 1(c) and 2(d)). The surface atoms with one dangling bond are less exposed than those with two dangling bonds. Figure 6(a) presents the percentage of the hydrogen atoms in surfaces (013) and (113) that bond to the most exposed carbon atoms. In general, ~76% of hydrogen atoms present in the (113) surface react with the most exposed carbon atoms and ~24% reach the atoms underneath them (the green ones in Figure 2(d)) until 900°C. As for the (013) surface, these numbers vary from 64% to 68%, in which 32% to 36% react with the carbon atoms with one dangling bond each (the green ones in Figure 1(c)) until 900°C. Above 900°C, the number of hydrogen atoms bonded to the most exposed carbon atoms decrease to 62% and 76% for (013) and (113), respectively. These results suggest that it is relatively difficult for hydrogen atoms to reach the less exposed carbon atoms with one dangling bond in the surface (013) than in (113), even with the efficient temperature-dependent passivation of the dangling bonds.

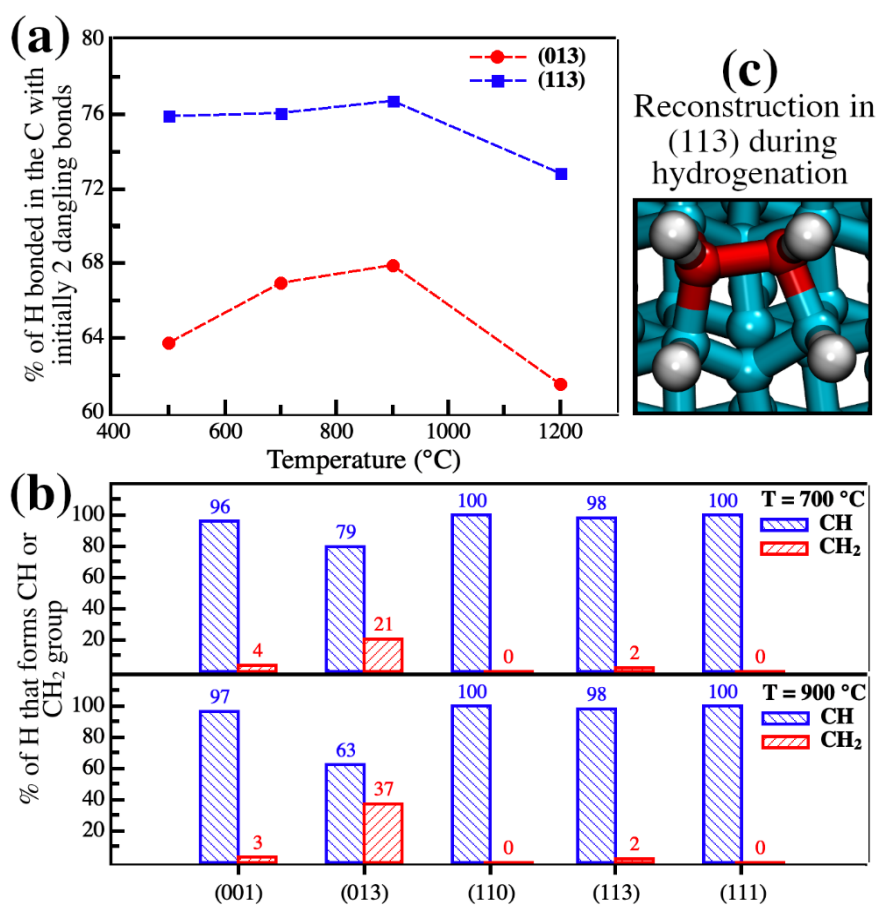

**Figure 6**. (a) Percentage of the hydrogen atoms that are bonded to the most exposed carbon atoms in the surfaces (013) and (113). (b) Percentage of the hydrogen atoms that form CH or $CH_2$ groups at the tested diamond surfaces. (c) A representation of the induced surface reconstructions during the hydrogenation of the (113) surface.

Beyond the formation of CH groups at the diamond surfaces, due to the presence of some defects and/or some carbon atoms with more than one dangling bond, it is also possible to form $CH_2$ and/or $CH_3$ groups. These other groups are detected in experiments [10], and they can have a direct influence on the electronic properties of hydrogenated diamond surfaces. According to theoretical results for hydrogenated Si and Ge surfaces [42], which have similar geometrical surfaces, the presence of $CH_2$ and/or $CH_3$ groups on the surface results in a decreased ionization potential, reduction in WF and a more negative EA in comparison to a surface with only CH groups. This result suggests that a similar effect can be expected for diamond surfaces. In Figure 6(b) we present the percentage of the hydrogen atoms incorporated into the diamond surface that form CH and $CH_2$ groups at 700°C and 900°C. As we are using pristine (non-defective) diamond surfaces, we did not observe the formation of the $CH_3$ group, since the exposed carbon atoms have just one or two dangling bonds. For the surfaces (110) and (111), we have only CH groups. This is expected since the exposed carbon atoms have only one dangling bond. However, for the other diamond surfaces, $CH_2$ groups are present, but in smaller quantities in (001) and (113) than in (013). As the (001) surface presents more than 70% of surface reconstruction at higher temperatures, we can expect to see a few $CH_2$ groups on it. As illustrated in Figure 6(b), the surface (001) allows 3%–4% of the incorporated hydrogen atoms in a $CH_2$ configuration between 700 and 900°C, which corresponds to ~1 $CH_2$ group for each 6 $nm^2$ of the surface.

We expected to observe a considerable number of $CH_2$ groups in (113) since it presents exposed carbon atoms with two dangling bonds. However, for both 700°C and 900°C, only 2% of the hydrogen atoms are in a $CH_2$ configuration, which represents ~1 $CH_2$ group for each 3 $nm^2$ of the surface. Further analysis showed that the hydrogenation process induces surface reconstructions involving the most exposed carbon atoms (highlighted in red in Figure 2(d)), as represented in Figure 6(c). These reconstructions decrease the number of dangling bonds and the probability of the $CH_2$ group formation. The induced reconstructions during hydrogenation can be understood from the geometrical configuration of the (113) surface. In this surface, the most exposed atoms (the red ones in Figure 2(d) that have two dangling bonds each) are at a distance of 2.5 Å from each other, but the distance between the hydrogen atoms bonded to these atoms increase significantly the electrostatic (repulsion) interactions. Then, it is energetically favorable that these atoms reconstruct and be bonded to only one hydrogen atom.

In the (013) surface, 21% and 37% of the incorporated hydrogen atoms form a $CH_2$ configuration at 700°C and 900°C, respectively. These results suggest that even the (013) surface is not the most efficient in the hydrogenation process (Figure 5(f)), the considerable number of $CH_2$ groups present at the surface may enhance the intrinsic electronic properties of the surface. To elucidate the role of these functional groups in modifying electronic properties, a detailed atomistic study is required. In contrast to the (113) surface, the most exposed atoms on (013) (the red ones in Figure 1(c) with two dangling bonds each) are a distant 3.1 Å from each other, which in turn reduces the probability of $CH_2$ group formation, as compared to the (113).

In summary, our results suggest that the surfaces (001), (110), and (113) are the most efficient in the hydrogenation process, but (113) could be the most attractive due to its additional lower electrical resistance compared to the other surfaces. If it is possible to produce diamond films with higher hydrogen coverage, this phase could be useful in producing new p-type materials with superior electronic properties for RF device applications.

4. Conclusions:

We have studied the surface characteristics and hydrogenation efficiency of five different bare diamond surfaces: (001), (013), (110), (113), and (111). Our results indicate that the number of dangling bonds at the surface is temperature-dependent for the (001) surface due to structural reconstructions, but it has a less pronounced effect on other surfaces. The (013) and (113) surfaces have the highest dangling bond densities of 29 and 30 dangling bonds per $nm^2$, respectively. These dangling bonds present the highest number of bonded hydrogen atoms at the surface. Quantitatively, these surfaces offer 16 and 17 hydrogen atoms per $nm^2$ as compared to the 12 and 14 per $nm^2$ offered by the (001) and (110) surfaces, respectively, at 1200°C. These surfaces, along with the (113) surface, exhibit the highest level of hydrogenation efficiencies, resulting in more than 60% of hydrogen coverage at high temperatures. However, the hydrogenation efficiencies for the surfaces (013) and (111) were relatively low. Regarding the hydrogenated groups formed at the surface, the CH group was favored (considering non-defective surfaces), but a considerable number of $CH_2$ groups were also formed at the surface (013). It is known that these groups can help in tuning the surface electron affinities and WF values, but further study is necessary to better address their effects.

To tune the electronic properties of the diamond surfaces to the desired negative electron affinities and decreased WFs, it is necessary to have a high number of hydrogen atoms and a small number of dangling bonds at the surface. Our results suggest that a diamond film with a high content of exposed (001), (110), and (113) surfaces could be beneficial to make hydrogenated diamond-based p-channel transistors. However, more studies regarding the electronic properties of these different diamond surfaces are still necessary to determine which condition is better: the highest number of hydrogen atoms or the lowest number of dangling bonds at the surface. We hope the present study can stimulate works along these lines.

5. Acknowledgments

EFO and DSG would like to thank the Brazilian agencies CNPq and FAPESP (Grants 2013/08293-7, 2016/18499-0, and 2019/07157-9) for financial support. Computational support from the Center for Computational Engineering and Sciences at Unicamp through the FAPESP/CEPID Grant No. 2013/08293-7 and the Center for Scientific Computing (NCC/GridUNESP) of São Paulo State University (UNESP) is also acknowledged. This work was based upon the ARL-Rice Initiative for an Integrative Program in Advanced Materials and Next-Generation Networks. (P.M.A., R.V., A.B.P., X.Z., C.L., M.R.N., P.B.S., A.G.B., T.G.I.).

6. <u>References:</u>

[1]  X. Liu, X. Chen, D.J. Singh, R.A. Stern, J. Wu, S. Petitgirard, C.R. Bina, S.D. Jacobsen, Boron–oxygen complex yields n-type surface layer in semiconducting diamond, Proc. Natl. Acad. Sci. 116 (2019) 7703–7711. https://doi.org/10.1073/pnas.1821612116.

[2]  T.T. Pham, N. Rouger, C. Masante, G. Chicot, F. Udrea, D. Eon, E. Gheeraert, J. Pernot, Deep depletion concept for diamond MOSFET, Appl. Phys. Lett. 111 (2017) 173503. https://doi.org/10.1063/1.4997975.

[3]  Y. Sasama, K. Komatsu, S. Moriyama, M. Imura, T. Teraji, K. Watanabe, T. Taniguchi, T. Uchihashi, Y. Takahide, High-mobility diamond field effect transistor with a monocrystalline h-BN gate dielectric, APL Mater. 6 (2018) 111105. https://doi.org/10.1063/1.5055812.

[4]  W. Fei, T. Bi, M. Iwataki, S. Imanishi, H. Kawarada, Oxidized Si terminated diamond and its MOSFET operation with $SiO_2$ gate insulator, Appl. Phys. Lett. 116 (2020) 212103. https://doi.org/10.1063/1.5143982.

[5]  C.J. Zhou, J.J. Wang, J.C. Guo, C. Yu, Z.Z. He, Q.B. Liu, X.D. Gao, S.J. Cai, Z.H. Feng, Radiofrequency performance of hydrogenated diamond MOSFETs with alumina, Appl. Phys. Lett. 114 (2019) 063501. https://doi.org/10.1063/1.5066052.

[6]  P.S. Mirabedini, B. Debnath, M.R. Neupane, P. Alex Greaney, A. Glen Birdwell, D. Ruzmetov, K.G. Crawford, P. Shah, J. Weil, Tony.G. Ivanov, Structural and electronic properties of 2D (graphene, hBN)/H-terminated diamond (100) heterostructures, Appl. Phys. Lett. 117 (2020) 121901. https://doi.org/10.1063/5.0020620.

[7]  H. Gomez, M.N. Groves, M.R. Neupane, Study of the structural phase transition in diamond (100) & (111) surfaces, Carbon Trends. 3 (2021) 100033. https://doi.org/10.1016/j.cartre.2021.100033.

[8]  S. Koizumi, ed., Power electronics device applications of diamond semiconductors, Woodhead Publishing, an imprint of Elsevier, Duxford, United Kingdom ; Cambridge, MA, United States, 2018.

[9]  H.O. Pierson, Handbook of carbon, graphite, diamond, and fullerenes: properties, processing, and applications, Noyes Publications, Park Ridge, N.J., U.S.A, 1993.

[10]  H. Kawarada, Hydrogen-terminated diamond surfaces and interfaces, Surf. Sci. Rep. 26 (1996) 205–206. https://doi.org/10.1016/S0167-5729(97)80002-7.

[11]  A.N. Andriotis, G. Mpourmpakis, E. Richter, M. Menon, Surface Conductivity of Hydrogenated Diamond Films, Phys. Rev. Lett. 100 (2008) 106801. https://doi.org/10.1103/PhysRevLett.100.106801.

[12] C. Sun, T. Hao, J. Li, H. Ye, C. Gu, The design and performance of hydrogen-terminated diamond metal-oxide-semiconductor field-effect transistors with high k oxide HfO2, Micro Nano Eng. 6 (2020) 100046. https://doi.org/10.1016/j.mne.2020.100046.

[13] P. Rivero, W. Shelton, V. Meunier, Surface properties of hydrogenated diamond in the presence of adsorbates: A hybrid functional DFT study, Carbon. 110 (2016) 469–479. https://doi.org/10.1016/j.carbon.2016.09.050.

[14] J. Liu, Y. Koide, Fabrication of Hydrogenated Diamond Metal–Insulator–Semiconductor Field-Effect Transistors, in: B. Prickril, A. Rasooly (Eds.), Biosens. Biodetection, Springer New York, New York, NY, 2017: pp. 217–232. https://doi.org/10.1007/978-1-4939-6911-1_15.

[15] F. Maier, J. Ristein, L. Ley, Electron affinity of plasma-hydrogenated and chemically oxidized diamond (100) surfaces, Phys. Rev. B. 64 (2001) 165411. https://doi.org/10.1103/PhysRevB.64.165411.

[16] J. van der Weide, Z. Zhang, P.K. Baumann, M.G. Wensell, J. Bernholc, R.J. Nemanich, Negative-electron-affinity effects on the diamond (100) surface, Phys. Rev. B. 50 (1994) 5803–5806. https://doi.org/10.1103/PhysRevB.50.5803.

[17] F. Maier, M. Riedel, B. Mantel, J. Ristein, L. Ley, Origin of Surface Conductivity in Diamond, Phys. Rev. Lett. 85 (2000) 3472–3475. https://doi.org/10.1103/PhysRevLett.85.3472.

[18] K.P.S.S. Hembram, S. Lee, H. Im, H. Ju, S.-H. Jeong, J.-K. Lee, The surface hybridization of diamond with vertical graphene: a new route to diamond electronics, Mater. Horiz. 7 (2020) 470–476. https://doi.org/10.1039/C9MH01588D.

[19] M. De La Pierre, M. Bruno, C. Manfredotti, F. Nestola, M. Prencipe, C. Manfredotti, The (100), (111) and (110) surfaces of diamond: an *ab initio* B3LYP study, Mol. Phys. 112 (2014) 1030–1039. https://doi.org/10.1080/00268976.2013.829250.

[20] S.J. Sque, R. Jones, P.R. Briddon, Hydrogenation and oxygenation of the (100) diamond surface and the consequences for transfer doping, Phys. Status Solidi A. 202 (2005) 2091–2097. https://doi.org/10.1002/pssa.200561911.

[21] A.C.T. van Duin, S. Dasgupta, F. Lorant, W.A. Goddard, ReaxFF: A Reactive Force Field for Hydrocarbons, J. Phys. Chem. A. 105 (2001) 9396–9409. https://doi.org/10.1021/jp004368u.

[22] S. Plimpton, Fast Parallel Algorithms for Short-Range Molecular Dynamics, J. Comput. Phys. 117 (1995) 1–19. https://doi.org/10.1006/jcph.1995.1039.

[23] K. Chenoweth, A.C.T. van Duin, W.A. Goddard, ReaxFF Reactive Force Field for Molecular Dynamics Simulations of Hydrocarbon Oxidation, J. Phys. Chem. A. 112 (2008) 1040–1053. https://doi.org/10.1021/jp709896w.

[24] I. Friel, S.L. Clewes, H.K. Dhillon, N. Perkins, D.J. Twitchen, G.A. Scarsbrook, Control of surface and bulk crystalline quality in single crystal diamond grown by chemical vapour deposition, Diam. Relat. Mater. 18 (2009) 808–815. https://doi.org/10.1016/j.diamond.2009.01.013.

[25] M.-A. Pinault-Thaury, S. Temgoua, R. Gillet, H. Bensalah, I. Stenger, F. Jomard, R. Issaoui, J. Barjon, Phosphorus-doped (113) CVD diamond: A breakthrough towards bipolar diamond devices, Appl. Phys. Lett. 114 (2019) 112106. https://doi.org/10.1063/1.5079924.

[26] V. Damle, K. Wu, O. De Luca, N. Ortí-Casañ, N. Norouzi, A. Morita, J. de Vries, H. Kaper, I.S. Zuhorn, U. Eisel, D.E.P. Vanpoucke, P. Rudolf, R. Schirhagl, Influence of diamond crystal orientation on the interaction with biological matter, Carbon. 162 (2020) 1–12. https://doi.org/10.1016/j.carbon.2020.01.115.

[27] E.F. Oliveira, A. Batagin-Neto, D.S. Galvao, On the sulfur doping of γ-graphdiyne: A Molecular Dynamics and DFT study, MRS Adv. 5 (2020) 2701–2706.

https://doi.org/10.1557/adv.2020.255.
[28] J. Joyner, E.F. Oliveira, H. Yamaguchi, K. Kato, S. Vinod, D.S. Galvao, D. Salpekar, S. Roy, U. Martinez, C.S. Tiwary, S. Ozden, P.M. Ajayan, Graphene Supported MoS2 Structures with High Defect Density for an Efficient HER Electrocatalysts, ACS Appl. Mater. Interfaces. 12 (2020) 12629–12638. https://doi.org/10.1021/acsami.9b17713.
[29] M.Z.S. Flores, P. a. S. Autreto, S.B. Legoas, D.S. Galvao, Graphene to graphane: a theoretical study, Nanotechnology. 20 (2009) 465704. https://doi.org/10.1088/0957-4484/20/46/465704.
[30] E. Marinho, P.A. da S. Autreto, Me-graphane: tailoring the structural and electronic properties of Me-graphene via hydrogenation, Phys. Chem. Chem. Phys. 23 (2021) 9483–9491. https://doi.org/10.1039/D0CP06684B.
[31] R. Paupitz, P. a. S. Autreto, S.B. Legoas, S.G. Srinivasan, A.C.T. van Duin, D.S. Galvão, Graphene to fluorographene and fluorographane: a theoretical study, Nanotechnology. 24 (2012) 035706. https://doi.org/10.1088/0957-4484/24/3/035706.
[32] F. Silva, J. Achard, X. Bonnin, O. Brinza, A. Michau, A. Secroun, K. De Corte, S. Felton, M. Newton, A. Gicquel, Single crystal CVD diamond growth strategy by the use of a 3D geometrical model: Growth on (113) oriented substrates, Diam. Relat. Mater. 17 (2008) 1067–1075. https://doi.org/10.1016/j.diamond.2008.01.006.
[33] Y. Sasama, K. Komatsu, S. Moriyama, M. Imura, S. Sugiura, T. Terashima, S. Uji, K. Watanabe, T. Taniguchi, T. Uchihashi, Y. Takahide, Quantum oscillations in diamond field-effect transistors with a h -BN gate dielectric, Phys. Rev. Mater. 3 (2019) 121601. https://doi.org/10.1103/PhysRevMaterials.3.121601.
[34] A.V. Petukhov, D. Passerone, F. Ercolessi, E. Tosatti, A. Fasolino, (Meta)stable reconstructions of the diamond (111) surface: Interplay between diamond and graphitelike bonding, Phys. Rev. B. 61 (2000) R10590–R10593. https://doi.org/10.1103/PhysRevB.61.R10590.
[35] A. Kromka, J. Čech, H. Kozak, A. Artemenko, T. Ižák, J. Čermák, B. Rezek, M. Černák, Low-Temperature hydrogenation of diamond nanoparticles using diffuse coplanar surface barrier discharge at atmospheric pressure: Low-temperature hydrogenation of diamond nanoparticles, Phys. Status Solidi B. 252 (2015) 2602–2607. https://doi.org/10.1002/pssb.201552232.
[36] V. Seshan, D. Ullien, A. Castellanos-Gomez, S. Sachdeva, D.H.K. Murthy, T.J. Savenije, H.A. Ahmad, T.S. Nunney, S.D. Janssens, K. Haenen, M. Nesládek, H.S.J. van der Zant, E.J.R. Sudhölter, L.C.P.M. de Smet, Hydrogen termination of CVD diamond films by high-temperature annealing at atmospheric pressure, J. Chem. Phys. 138 (2013) 234707. https://doi.org/10.1063/1.4810866.
[37] Sh. Michaelson, R. Akhvlediani, A. Hoffman, Hydrogenation and thermal stability of nano- and microcrystalline diamond films studied by vibrational electron spectroscopy, J. Appl. Phys. 104 (2008) 083527. https://doi.org/10.1063/1.3000662.
[38] B. Rezek, C.E. Nebel, Electronic properties of plasma hydrogenated diamond surfaces: A microscopic study, Diam. Relat. Mater. 15 (2006) 1374–1377. https://doi.org/10.1016/j.diamond.2005.10.002.
[39] L. Ostrovskaya, V. Perevertailo, V. Ralchenko, A. Dementjev, O. Loginova, Wettability and surface energy of oxidized and hydrogen plasma-treated diamond films, Diam. Relat. Mater. 11 (2002) 845–850. https://doi.org/10.1016/S0925-9635(01)00636-7.
[40] J.-C. Arnault, T. Petit, H. Girard, A. Chavanne, C. Gesset, M. Sennour, M. Chaigneau, Surface chemical modifications and surface reactivity of nanodiamonds hydrogenated by CVD plasma, Phys. Chem. Chem. Phys. 13 (2011) 11481.

https://doi.org/10.1039/c1cp20109c.

[41] W.F. Paxton, S. Ravipati, M.M. Brooks, M. Howell, J.L. Davidson, Thermionic Emission from Diamond Films in Molecular Hydrogen Environments, Front. Mech. Eng. 3 (2017) 18. https://doi.org/10.3389/fmech.2017.00018.

[42] I. Marri, M. Amato, M. Bertocchi, A. Ferretti, D. Varsano, S. Ossicini, Surface chemistry effects on work function, ionization potential and electronic affinity of Si(100), Ge(100) surfaces and SiGe heterostructures, Phys. Chem. Chem. Phys. 22 (2020) 25593–25605. https://doi.org/10.1039/D0CP04013D.